\documentclass[letterpaper,11pt]{article}
\usepackage{graphicx} 
\usepackage[paperwidth=8.5in, paperheight=11.80in,margin=1in]{geometry}
\usepackage{setspace}
\usepackage{tikz}
\usepackage[strict]{changepage}
\usepackage{amsthm}
\usepackage{float}
\usepackage{fancyhdr}
 \usepackage{setspace}
 \usepackage{amssymb}
 \usepackage{parskip}
 \usepackage{tabularx, booktabs}
 \usepackage{amsmath}
 \usepackage{graphicx}
\theoremstyle{definition}

\usepackage{authblk}
\usetikzlibrary{shapes, arrows.meta, positioning}
\usepackage{tcolorbox}
\usepackage{subcaption} 
\usepackage[utf8]{inputenc}
\usepackage{csquotes}

\usepackage[colorlinks=true, linkcolor=blue, citecolor=blue, urlcolor=blue]{hyperref}

\usepackage[
  backend=biber,
  bibstyle=apa,          
  citestyle=numeric-comp,
  sorting=none
]{biblatex}

\DeclareLanguageMapping{american}{american-apa}
\DeclareFieldFormat{labelnumberwidth}{\mkbibbrackets{#1}}
\defbibenvironment{bibliography}
  {\list
     {\printfield[labelnumberwidth]{labelnumber}}
     {\setlength{\labelwidth}{\labelnumberwidth}%
      \setlength{\leftmargin}{\labelwidth}%
      \setlength{\labelsep}{\biblabelsep}%
      \addtolength{\leftmargin}{\labelsep}%
      \setlength{\itemsep}{\bibitemsep}%
      \setlength{\parsep}{\bibparsep}}}
  {\endlist}
  {\item}

\begin{document}

\title{\textbf{Bayesian Hierarchical Methods for Surveillance of Cervical Dystonia Treatments}}
\author[1*]{Dennis Baidoo}
\author[1]{Emmanuel Kubuafor}
\author[2]{Samuel Frimpong Osarfo}
\author[3]{Frank Amo Agyei-Owusu}
\author[4]{Justice Akuoko Frimpong}
\author[5]{Robert Amevor}
\author[6]{Agnes Duah}
\author[7]{Fuseini Aboagye}

\affil[1]{Department of Mathematics and Statistics, University of New Mexico, Albuquerque, USA}
\affil[2]{Department of Statistics, Oklahoma State University, Stillwater, OK, USA}
\affil[3]{School of Public Health, Virginia Commonwealth University, VA, USA}
\affil[4]{Department of Biostatistics, University of Michigan, Ann Arbor, MI, USA}
\affil[5]{Arnold School of Public Health, University of South Carolina, Columbia, SC, USA}
\affil[6]{School of Public Health-Bloomington, Indiana University, IN, USA}
\affil[7]{Department of Statistics and Actuarial Science, University of Ghana, Accra}
\affil[ ]{*\texttt{Corresponding author: baidennwin@gmail.com}}
\date{}

\maketitle
\onehalfspacing
    
\section*{Abstract}

\textbf{Background}: Cervical dystonia, a debilitating neurological disorder marked by involuntary muscle contractions and chronic pain, presents significant treatment challenges despite advances in botulinum toxin therapy. While botulinum toxin type B has emerged as one of the leading treatments, comparative efficacy across doses and the influence of demographic factors for personalized medicine remain understudied. \textbf{Objectives}: This study aimed to: (1) compare the efficacy of different botulinum toxin type B doses using Bayesian methods, (2) evaluate demographic and clinical factors affecting treatment response, and (3) establish a probabilistic framework for personalized cervical dystonia management. \textbf{Methods}: We analyzed data from a multicenter randomized controlled trial involving 109 patients assigned to placebo, 5,000 units, or 10,000 units of botulinum toxin type B groups. The primary outcome was the Toronto Western Spasmodic Torticollis Rating Scale measured over 16 weeks. Bayesian hierarchical modeling assessed treatment effects while accounting for patient heterogeneity. \textbf{Results}: Lower botulinum toxin type B doses (5,000 units) showed greater overall Toronto Western Spasmodic Torticollis Rating Scale score reductions (treatment effect: -2.39, 95\% credible interval: -4.10 to -0.70) over 10,000 units doses. Male patients demonstrated better responses (5.2\% greater improvement) than female patients. Substantial between-patient variability and site-specific effects were observed, highlighting the need for personalized protocols. 
\textbf{Conclusions}: The study confirms botulinum toxin type B's dose-dependent efficacy while identifying key modifiable factors in treatment response. Bayesian methods provided nuanced insights into uncertainty and heterogeneity, revolutionizing the understanding of botulinum toxin type B pharmacodynamics and paving the way for personalized medicine in cervical dystonia management.

\textbf{Keywords}: Cervical dystonia, Bayesian hierarchical modeling, Botulinum toxin type B, Personalized  medicine, Clinical trial analysis

\newpage
\section{Introduction}
     Cervical dystonia (CD) is a complex neurological disorder characterized by involuntary head and neck movements, often accompanied by pain and non-motor symptoms. The pathophysiology of CD remains unclear, but genetic factors, trauma, sensory system involvement, and basal ganglia dysfunction may contribute to its development \cite{sawek_2021_botulinum,stacy_2000_idiopathic,newby_2017_a}. Most unsatisfactory responses to botulinum toxin treatment (BotB) in cervical dystonia patients resulted from correctable factors like suboptimal dosing or muscle targeting, with 78\% achieving satisfactory outcomes after treatment adjustments \cite{jinnah_2016_botulinum}.
    Recent research has revealed whole-brain white matter abnormalities in CD patients \cite{zito_2023_fixelbased}. Treatment options include oral medications, such as anticholinergics and GABAmimetic agents, with limited efficacy, and BotB injections have revolutionized CD treatment, offering high response rates with minimal side effects. For patients resistant to botulinum toxin or oral medications, surgical interventions like selective dorsal ramisectomy or deep-brain stimulation may be considered \cite{sawek_2021_botulinum,adler_2000_pharmacological} . A comparative study of various botulinum toxin formulations, Dysport, Botox, Xeomin, and Myobloc, showed no meaningful differences in effectiveness four weeks after injection \cite{han_2014_a}. Bayesian hierarchical methods offer powerful tools for analyzing surveillance data in public health. They enable probabilistic assessment of outbreak intensity, incorporating clinical factors, potentially providing earlier detection, and also yield findings that are more meaningful for clinical practice compared to conventional frequentist approaches, as they directly quantify the likelihood that a particular treatment is optimal or ranks among the most effective options \cite{buenconsejo_2008_a,chan_2010_probabilistic,carlin_2013_case,christensen_2011_bayesian}. For conditions such as cervical dystonia, Bayesian hierarchical approaches may enhance monitoring capabilities by quantifying uncertainties in decision-making and synthesizing varied clinical data, leading to better recommendations and population-level analysis \cite{asimakidou_2023_a,edwards_2012_a}.
    This study aims to utilize Bayesian hierarchical modeling to assess treatment effects in CD, focusing on comparing the efficacy of different BotB formulations and exploring the avenue for personalized treatment intervention for patients. By addressing these key gaps, the study strives to support the development of personalized treatment approaches, improve patient outcomes, and establish a foundation for future research.
\begin{figure}[h]
\centering
\scalebox{0.9}{ 
\begin{tikzpicture}[
    node distance=1.5cm,
    box/.style={rectangle, draw, rounded corners, text width=4cm, minimum height=1cm, align=center},
    bigbox/.style={rectangle, draw, rounded corners, text width=5.5cm, minimum height=1.8cm, align=center},
    oval/.style={ellipse, draw, text width=3cm, minimum height=1cm, align=center},
    arrow/.style={-Stealth, thick}
]

\node[box, fill=blue!10] (problem) {Clinical Problem: \\ Cervical Dystonia Treatment Variability};
\node[bigbox, fill=green!10, below=of problem] (data) {Multicenter RCT Data \\ (N=109 patients) \\ 3 Treatment Arms: \\ Placebo, 5000U, 10000U BotB};
\node[box, fill=red!10, right=of data] (model) {Bayesian Hierarchical Model \\ (Dose, Sex, Age, Clinical Site effects, etc.)};
\node[box, fill=purple!10, below=of model] (results) {Key Findings: \\ - Dose-response gradients \\ - Sex/site variability \\ - Temporal patterns};
\node[oval, fill=yellow!10, below=of results] (impact) {Clinical Impact: \\ Personalized Treatment Protocols};

\draw[arrow] (problem) -- (data);
\draw[arrow] (data) -- (model);
\draw[arrow] (model) -- (results);
\draw[arrow] (results) -- (impact);

\end{tikzpicture}
}
\caption{Conceptual framework of the Bayesian analysis of botulinum toxin type B efficacy in cervical dystonia treatment.}
\label{fig:framework}
\end{figure}
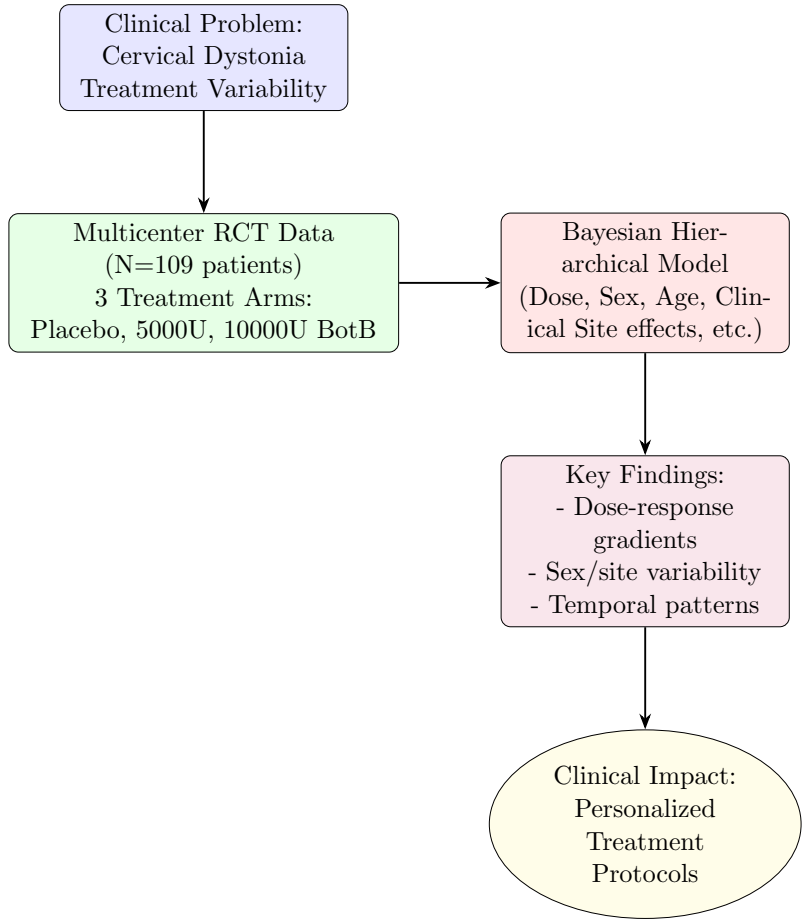

\section{Methods}
\subsection{Data description and framework}
The secondary source of the dataset used in the study is derived from a multicenter, randomized controlled trial to evaluate the efficacy of BotB in treating CD. It includes longitudinal data collected from nine different clinical sites in the United States of America. The study design encompasses randomization into three groups: placebo (N=36), 5,000 units dosing of BotB (N=36), and 10,000 units dosing of BotB (N=37) summarized in Figure 2A. The primary response variable is the total score on the Toronto Western Spasmodic Torticollis Rating Scale (TWSTRS), a composite measure of severity, pain, and disability associated with CD. Measurements were conducted at baseline (week 0) and 2, 4, 8, 12, and 16 weeks following treatment initiation. The dataset consists of 631 observations with associated covariates, including week, clinical site or location, treatments assigned, age of the patient, the biological sex of the patient, and total TWSTRS score, where higher scores indicate greater impairment for all 109 patients \cite{davis_2002_statistical, harrelljr_2003_cervical, boyce_2012_active,comella_2015_reliability}. Figure 1 presents a step-by-step analysis of botulinum toxin treatment for CD using Bayesian methods.  Furthermore, it shows how these results can guide personalized treatment plans, connecting data analysis to real-world clinical decisions.

\begin{figure}[h]
\centering

\begin{subfigure}{0.49\textwidth}
    \centering
    \caption*{(A)}
    \includegraphics[width=\linewidth]{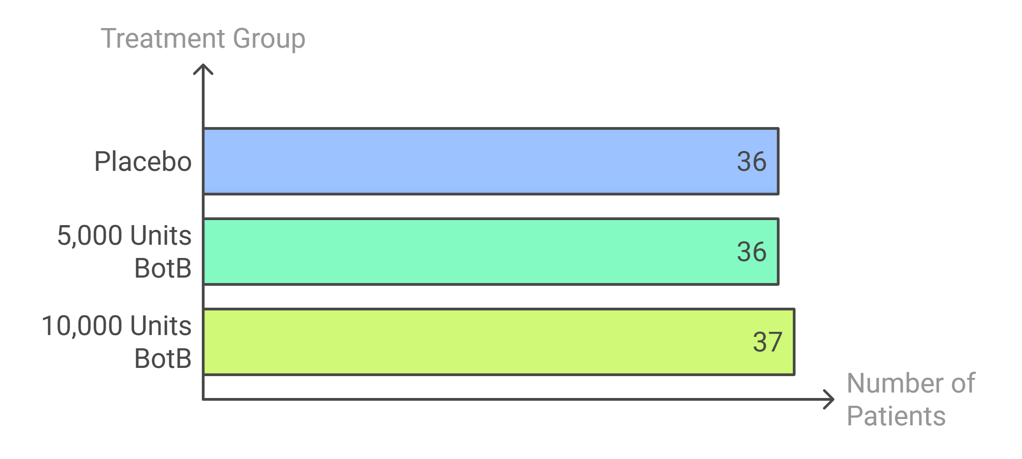}
    \label{fig:Bayes3}
\end{subfigure}
\hfill
\begin{subfigure}{0.46\textwidth}
    \centering
    \caption*{(B)}
    \includegraphics[width=\linewidth]{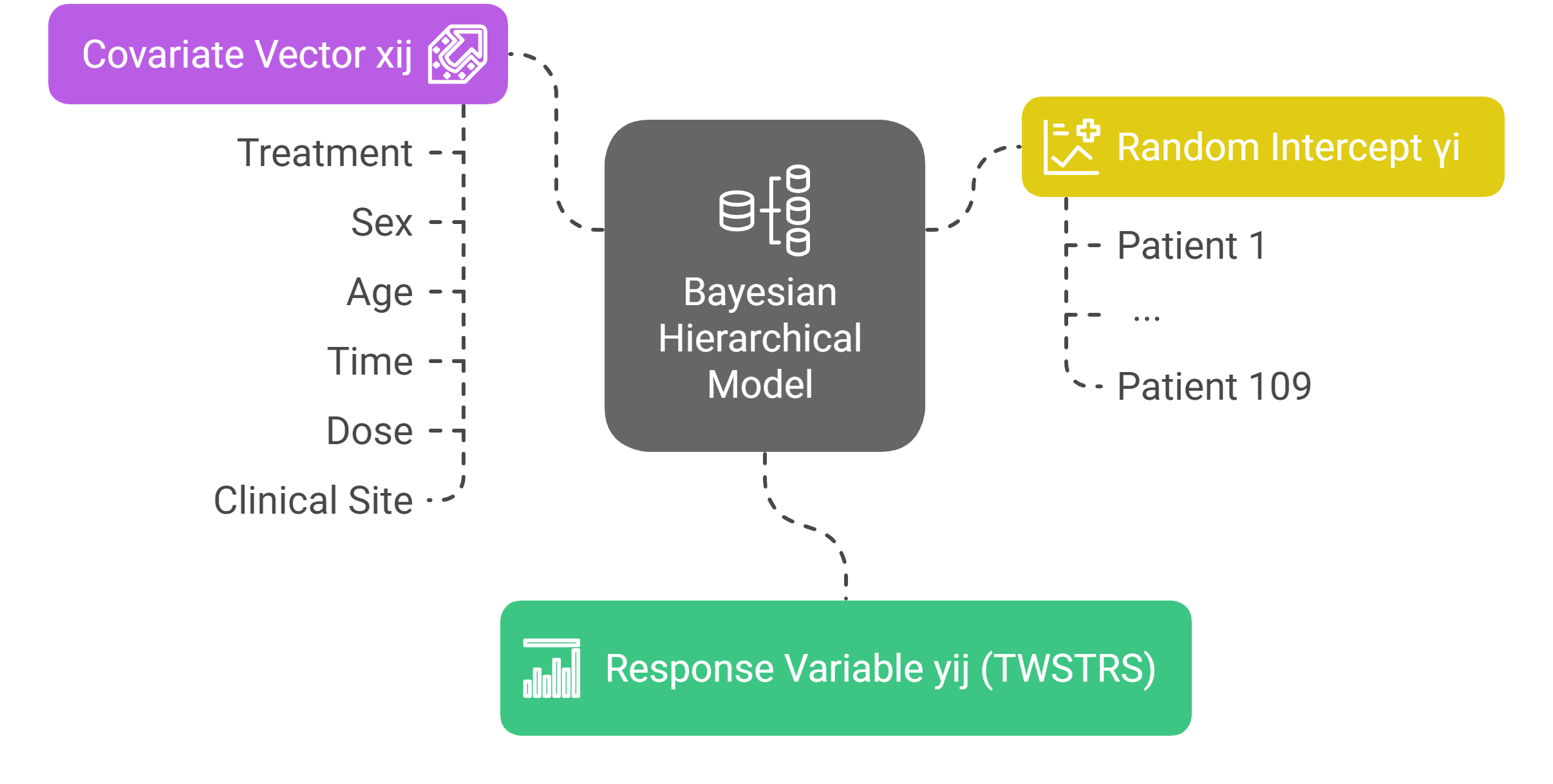}
    \label{fig:Bayes4}
\end{subfigure}

\caption{ 
(A) Distribution of patients across treatment groups,  
(B) Hierarchical model structure with covariates and random intercepts.}
\label{fig:BayesAB}
\end{figure}

\subsection{Bayesian hierarchical model building}
This section focuses on the statistical methods of the longitudinal data using a random intercept model (equation 1and Figure 2B), which is suitable for handling correlated measurement structures over time. This model is ideally suited for our CD surveillance study, as it enables probabilistic evaluation of treatment efficacy (e.g., BotB dose-response relationships) while quantifying uncertainty in motor outcomes. The approach allows us to: 1.) estimate posterior probabilities of treatment superiority across clinical sites, 2.) model temporal patterns of symptom relief and recurrence, and 3.) account for patient-level heterogeneity in the longitudinal TWSTRS data. By incorporating prior evidence on BotB effects, the model efficiently synthesizes multicenter trial data to optimize personalized treatment strategies. The random intercept model is a mixed model that accounts for both the inter-variability between patients and the intra-variability within a patient \cite{christensen_2011_bayesian}, as shown in equation (1) and Figure 2B.

\newpage
\noindent \textbf{Covariance structure:}

\begin{align}
  Cov(y_{ij}, y_{i^\prime j^\prime}) = \begin{cases}
\sigma_{score}^2 + \sigma_\gamma^2, & (i, j) = (i^\prime, j^\prime) \\
\sigma_\gamma^2, & i = i^\prime \text{ and } j \neq j^\prime \\
0, & i \neq i^\prime
\end{cases} 
\end{align}

The model, represented by equation (2), includes a response variable $y_{ij}$ (TWSTRS) and a covariate vector $x_{ij}$ recorded at time $j$ for each patient $i$ (Figure 2B). The covariate vector comprises treatment, sex, age, time (week-linear component), time squared (week*week, non-linear component), dose (time-adjusted dose variable accounting for the 2-week delay in treatment effect), and clinical site. The random intercept $\gamma_i$ varies for each patient in the study (equation 1).

\noindent \textbf{Hierarchical prior structures:}
\begin{align}
y_{ij} &= \gamma_i + x_{jk}^\prime\beta_k + \epsilon_{ij} \\
y_{ij} | \mu_j, \tau_{\text{score}} &\sim \text{Normal}(\mu_j, \tau_{\text{score}} = \sigma^{-2}_{\text{score}}) \\
\mu_j | \mu_{\beta}, \tau_{\gamma} &\sim \text{Normal}(\mu_{\beta}, \tau_{\gamma} = \sigma^{-2}_{\gamma}) \\
\epsilon_{ij} | \tau_{\text{score}} &\sim \text{Normal}(0, \tau_{\text{score}} = \sigma^{-2}_{\text{score}})
\end{align}

\noindent \textbf{Prior distributions:}

\begin{align}
P(\mu_{\beta}) &\sim \text{Normal}(a_{\mu_{\beta}} = 0, b_{\mu_{\beta}} = 0.000001) \\
P(\tau_{\gamma}) &\sim \text{Uniform}(a_{\sigma_{\gamma}} = 0, b_{\sigma_{\gamma}} = 1000) \\
P(\tau_{\beta}) &\sim \text{Gamma}(a_{\tau_{\beta}} = 0.001, b_{\tau_{\beta}} = 0.001); \tau_{\beta} = \sigma^{-2}_{\beta} \\
P(\tau_{\text{score}}) &\sim \text{Uniform}(0, 1000)
\end{align}

\noindent \textbf{Joint prior:}
\begin{align}
P(\mu_{\beta}, \tau_{\text{score}}, \tau_{\gamma}, \tau_{\beta}) &= P(\mu_{\beta})P(\tau_{\text{score}})P(\tau_{\gamma})P(\tau_{\beta})
\end{align}

\noindent \textbf{Posterior distribution:}
\begin{align}
P(\mu_{\beta}, \tau_{\text{score}}, \tau_{\gamma}, \tau_{\beta} | y) &\propto P(y | \mu_{\beta}, \tau_{\text{score}}, \tau_{\gamma}, \tau_{\beta}) P(\mu_{\beta})P(\tau_{\text{score}})P(\tau_{\gamma})P(\tau_{\beta})
\end{align}

Where $i = 1, . . . , 109$ patients, $j = 1, . . . , 631$ observations
and $k = 1, . . . , 15$ beta coefficients. 
 Equations (6)-(9) specify the likelihood, while (2)-(5) define the hierarchical prior structures, and equations (10) and (11) specify the prior joint and posterior distribution, respectively. These structures allow information to be shared across patients while estimating treatment effects. The model-building process will validate and determine the covariates to be included in the final model \cite{buenconsejo_2008_a,asimakidou_2023_a,christensen_2011_bayesian}.

\section{Results of Analysis}
In this section, we analyzed the raw data using descriptive statistics and validated the descriptive analysis results using our proposed Bayesian statistical model. This approach combines initial data exploration with model-based validation to ensure thorough, reliable conclusions.
\subsection{Comparative analysis of baseline variables}
The density plot (Figure 3A) shows the distribution of TWSTRS scores across the three treatment groups: Placebo, 5,000 units dosing, and 10,000 units dosing. The 10,000 units dosing group demonstrates a slight rightward shift, indicating relatively higher TWSTRS scores and poor severity reduction in pain improvement in CD symptoms compared to the Placebo and 5000 units dosing groups, which have overlapping distributions around higher scores. Also, the distribution of TWSTRS scores by sex (Figure 3B), females exhibit a slightly higher concentration of TWSTRS scores around the mid-range compared to males. This suggests that male patients respond better to the BotB treatments compared to female patients.

\begin{figure}[h]
    \centering
    \includegraphics[width=0.45\textwidth]{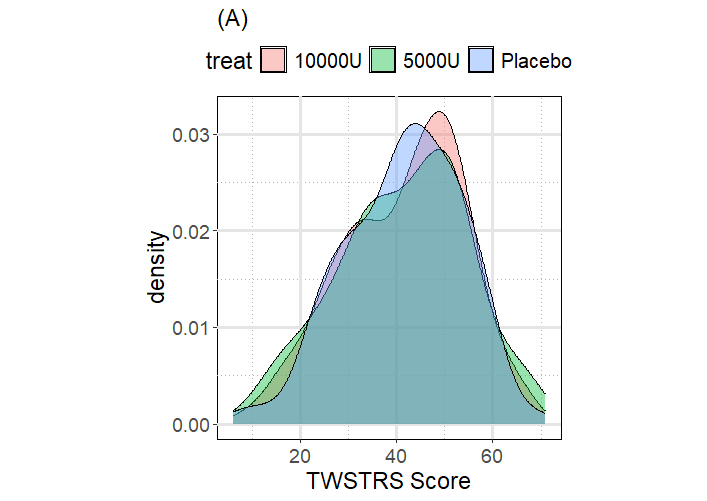} 
 \hspace{-14mm} 
    \includegraphics[width=0.45\textwidth]{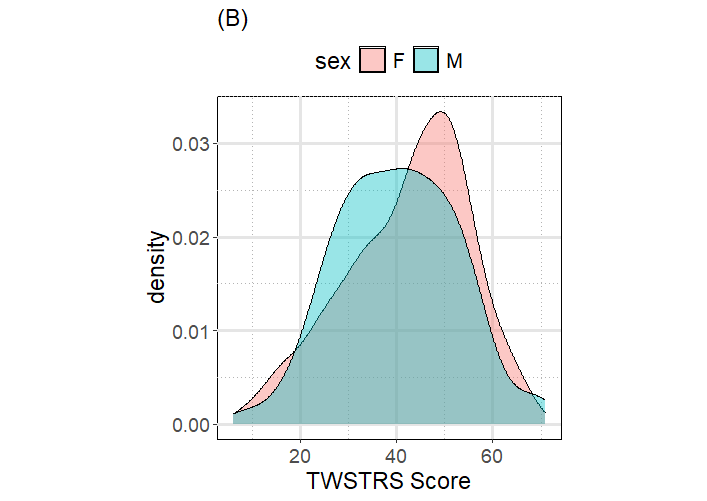} 
    \label{fig:score_distributionsA}
    \includegraphics[width=0.45\textwidth]{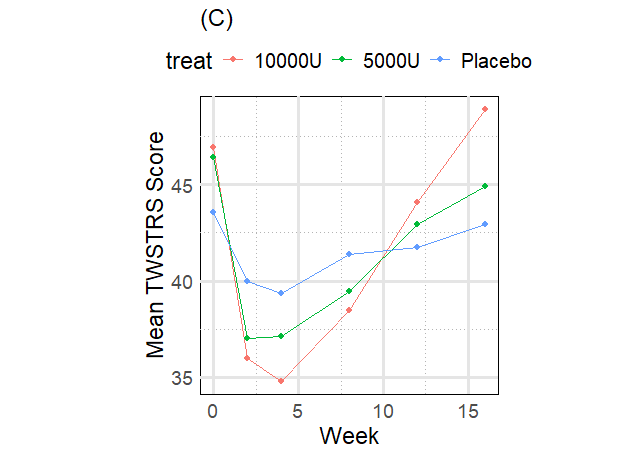} 
\hspace{-14mm} 
    \includegraphics[width=0.45\textwidth]{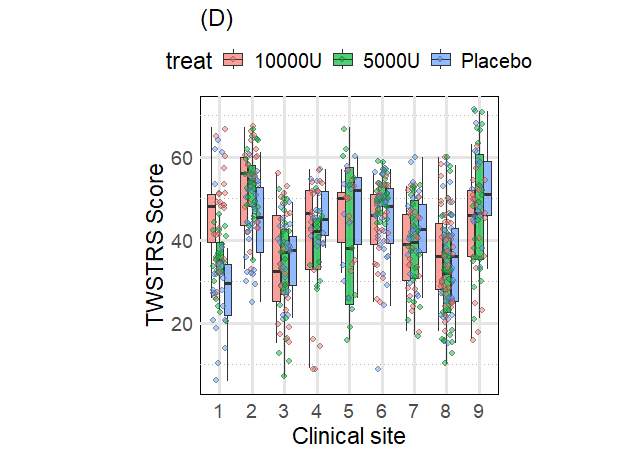} 
    \caption{(A) Density distribution of TWSTRS scores across treatment groups; (B) Score distribution by sex; (C) Mean TWSTRS score trajectories over time; (D) Score variability by clinical site}
    \label{fig:score_distributionsB}

\end{figure}

The interaction plot (Figure 3C) shows the three treatment groups' mean TWSTRS scores over time. The placebo dosing group demonstrates the most significant reduction in scores across all trial weeks, followed by the 5,000 unit dosing group, and finally by the 10,000 unit dosing group, with all treatment arms showing a gradual increase in TWSTRS scores after week 5. The treatment 10,000 units dosing group shows a sharper negative slope initially, transitioning to a positive slope after week 4, compared to the relatively flatter trends of the other groups. There is a curvilinear trend in mean TWSTRS scores across all treatment groups by weeks, with initial declines followed by gradual increases over time (Figure 3C). In addition, the box plot (Figure 3D) displays the distribution of TWSTRS scores across nine different clinical locations in the United States of America for the three treatment groups. Variability in TWSTRS scores is evident within and across clinical locations, with some treatments showing consistently lower median TWSTRS scores compared to others.

\begin{figure}[h]
\centering

\begin{subfigure}{0.4\textwidth}
    \centering
    \caption*{}
    \includegraphics[width=\linewidth]{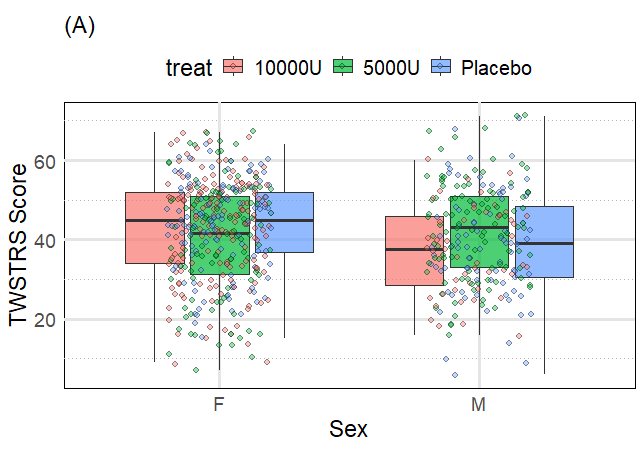}
    \label{fig:AAA}
\end{subfigure}
\hfill
\begin{subfigure}{0.4\textwidth}
    \centering
    \caption*{}
    \includegraphics[width=\linewidth]{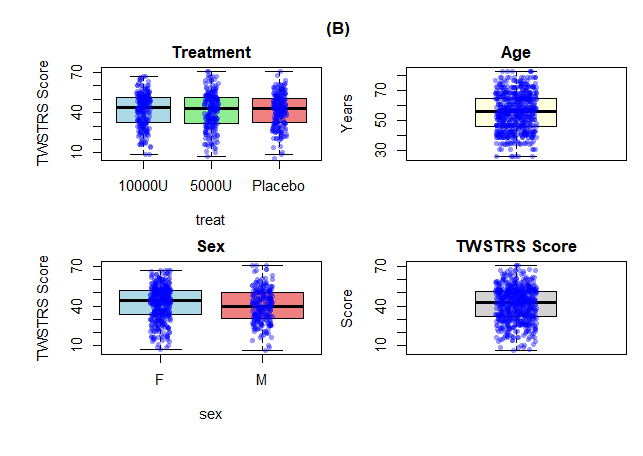}
    \label{fig:BBB}
\end{subfigure}
\hfill
\begin{subfigure}{0.4\textwidth}
    \centering
    \caption*{(C)}
    \includegraphics[width=\linewidth]{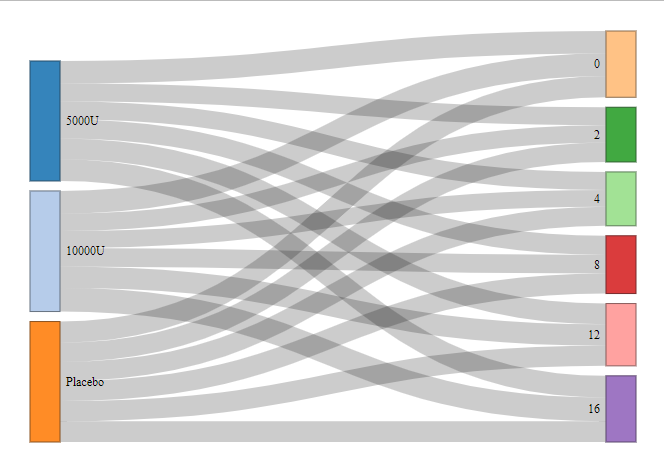}
    \label{fig:CCC}
\end{subfigure}
\hfill
\begin{subfigure}{0.45\textwidth}
    \centering
    \caption*{(D)}
    \includegraphics[width=\linewidth]{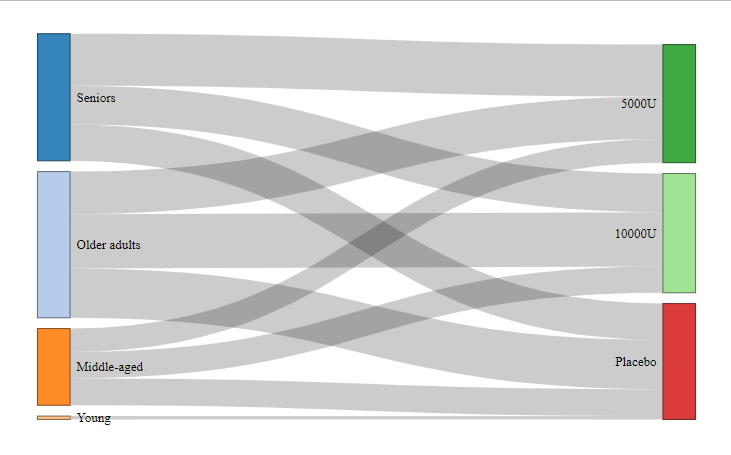}
    \label{fig:DDD}
\end{subfigure}
\caption{(A) Distribution of TWSTRS scores by sex and treatment group; (B) Boxplots of TWSTRS
scores by treatment, sex, and age; (C) Sankey diagram of treatment allocation and clinical outcome; (D) Patient flow across age groups and treatment categories.}
\label{fig:all}

\end{figure}

Placebo, 5000 units dosing, and 10000 units dosing were approximately equally assigned and common in seniors (aged $>$ 60), older adults (aged 45-60), and middle-aged (aged 30-45) but young patients (aged $<$ 30) mainly received placebo accounting for the smallest age group based on treatment randomization, as shown by the thin placebo-young flow line (Figure 4D). The distribution of TWSTRS scores by sex and treatment group (Figure 4A) shows that, on average, female patients generally had slightly higher TWSTRS scores than males across all treatments, with similar spreads in the data as indicated by the box sizes and whiskers. All three treatment groups appear to have overlapping distributions within each sex category.

The Sankey diagram (Figure 4C) shows how patients moved through a clinical trial over 16 weeks, comparing three treatment groups (5000 units dosing, 10000 units dosing, and Placebo). The gray flowing lines track patients from their initial treatment groups to different time points, with thicker lines representing more patients and vice versa. The visualization reveals similar retention patterns across all treatment groups, with consistent patient participation from baseline through the various time points (0, 2, 4, 8, 12, and 16 weeks), suggesting good overall adherence to the study protocol. The set of box plots (Figure 4B) summarizes the distributions of variables in the cervical dystonia dataset, including treatment group, age, sex, and TWSTRS scores. Notable patterns include lower TWSTRS scores in the 5,000 units dosing group and minimal differences between sexes. 
In addition, patients' average age and TWSTRS scores varied across clinical sites, with some sites enrolling older patients and others showing higher baseline severity. Despite this, the timing of assessments was consistent, averaging around week 7 across all clinical sites.

Figure 5A shows how different covariates in the study relate to each other. An association was observed between time (week) and symptom severity (TWSTRS score), with a correlation of 0.104. In addition, a noticeable relationship was identified between patient sex and treatment type. Other factors, like age and clinical location, showed weak relationships. Overall, while some small patterns exist, no strong connections were found between patient characteristics and treatment outcomes. These results indicate that while symptoms vary modestly over time, patient factors considered jointly contribute little to predicting treatment response, underscoring the importance of modeling potential interactions since clinical factors may not act independently.
The baseline characteristics showed comparable severity across treatment groups, with mean TWSTRS scores of 41.56 (SD = 12.55) for the 10,000-unit dosing group, 41.36 (SD = 13.53) for the 5,000-unit dosing group, and 41.51 (SD = 12.08) for placebo. Similarly, participants demonstrated balanced demographic profiles, with females showing mean TWSTRS scores of 42.26 and males 40.17, while average ages were nearly identical (55.63 vs. 55.60 years, respectively). These closely matched baseline measurements support the successful randomization of the study population and confirm initial equivalence between treatment arms \cite{berger_2004_ensuring,altman_1985_comparability}.

\subsection{Bayesian hierarchical model implementation}
The model-building and selection process for complex models, such as the random intercept model, relies on examining the beta densities of each variable generated through Markov chain Monte Carlo (MCMC) sampling in R, using the coda and rjags packages. R was solely used for these analyses. The model is then iteratively reduced by removing one insignificant variable at a time and re-running the MCMC sampling.
The final model is obtained when all variables are significant or backward selection has been exhausted. The posterior sigma distributions of the full and reduced models will be compared using density plots to validate the removal of an insignificant covariate. These processes will build systematic and sound statistical procedures in choosing the final model in explaining the variability in TWSTRS scores.

\subsection{Bayesian hierarchical model inference}
The analysis examined the influence of clinical and demographic factors on CD severity as measured by the TWSTRS total score. A Bayesian regression model (equation 1-11) was fitted with covariates including treatment, age, week, week², sex, dose, clinical site, and relevant interaction terms. 

\begin{figure}[h]
\centering

\begin{subfigure}{0.4\textwidth}
    \centering
    \includegraphics[width=\linewidth]{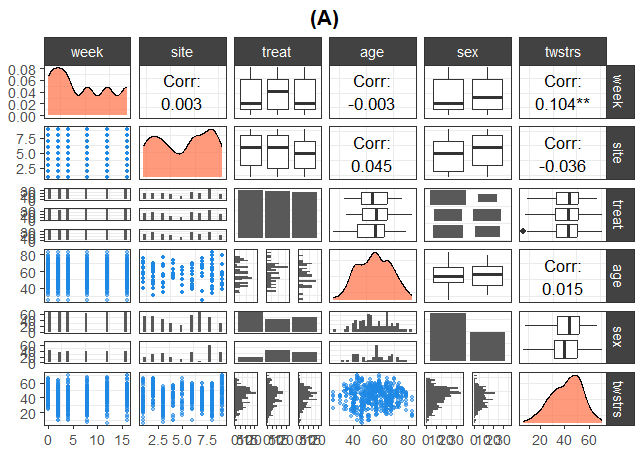}
    \label{fig:PPP}
\end{subfigure}
\hfill
\begin{subfigure}{0.45\textwidth}
    \centering
    \includegraphics[width=\linewidth]{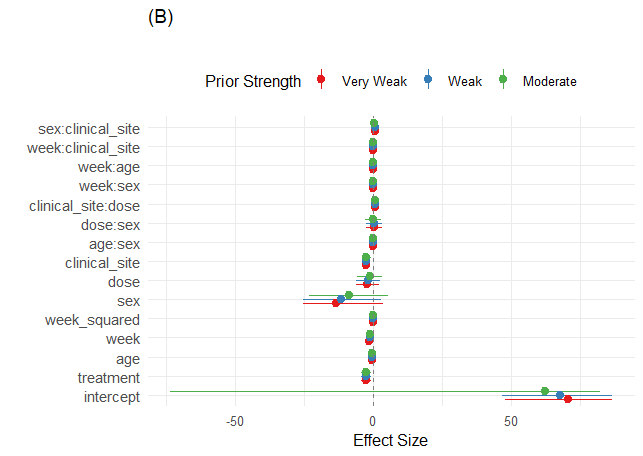}
    \label{fig:FFF}
\end{subfigure}

\vspace{0.5cm} 

\begin{subfigure}{0.4\textwidth}
    \centering
    \caption*{(C)}
    \includegraphics[width=\linewidth]{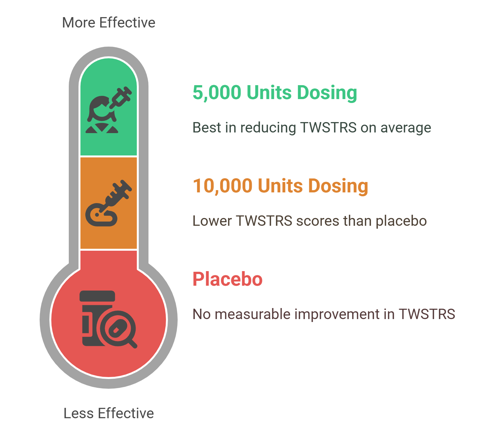}
    \label{fig:III}
\end{subfigure}
\hfill
\begin{subfigure}{0.45\textwidth}
    \centering
     \caption*{(D)}
     \vspace{11mm}
    \includegraphics[width=\linewidth]{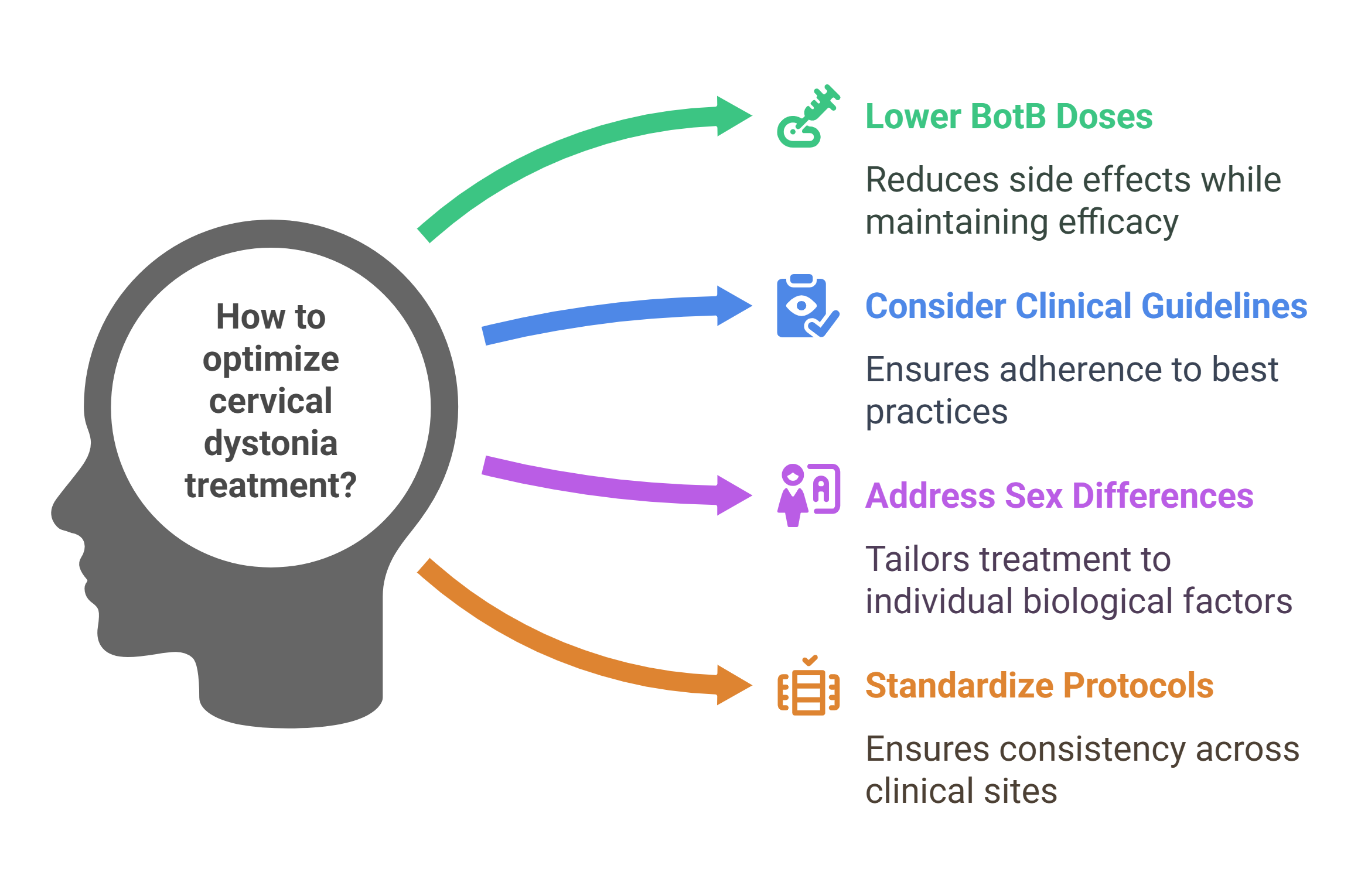}
    \label{fig:EEE}
\end{subfigure}

\caption{(A) Patient characteristics association summaries, (B) Full model Prior sensitivity analysis, (C) BotB treatments effect on TWSTRS, and (D) Clinical recommendation.}
\label{fig:GGG}
\end{figure}

The density plots for the posterior distribution for the full model for all parameters revealed the range of possible values for each regression coefficient and the likelihood of each value, and the significance of each covariate was evaluated by assessing whether the 95\% credible interval (CI) of its regression coefficient included zero. Coefficients with intervals excluding zero were considered statistically significant, indicating meaningful associations with TWSTRS scores. This approach allowed for rigorous evaluation of both main effects and interaction terms, identifying which covariates, such as treatment duration, dose adjustments, clinical site effects, or sex-specific responses, significantly influenced clinical outcomes. The results provide insights into modifiable and non-modifiable covariates of treatment response in CD.

We conducted sensitivity analyses using three prior strength settings: 
very weak ($\tau_{\beta} = 0.00001$, $\tau_{\text{score}} = 0.001$, $\tau_{\gamma} = 0.001$), 
weak ($\tau_{\beta} = 0.0001$, $\tau_{\text{score}} = 0.01$, $\tau_{\gamma} = 0.01$), 
and moderate ($\tau_{\beta} = 0.001$, $\tau_{\text{score}} = 0.1$, $\tau_{\gamma} = 0.1$). 
Graphically, the plot in Figure 5B presents the sensitivity analysis comparing how different prior assumptions (very weak, weak, moderate) affect the estimated effects of clinical and demographic factors on treatment outcomes. The key findings reveal that most main effects - including treatment, sex, age, dose, trial duration, and clinical site show consistent effect sizes across all prior specifications. However, the intercept exhibits moderate variability between prior settings, suggesting these findings are slightly sensitive to prior beliefs. This analysis confirms that while core treatment effects and primary demographic relationships remain stable, intercept effects require more cautious interpretation due to their slight dependence on prior specifications. Results remained consistent across all specifications, demonstrating the robustness of our primary findings to different prior assumptions. 
The treatment effects and key demographic relationships showed minimal variation ($<5\%$ change in effect sizes) when comparing the most conservative (very weak) and more informative (moderate) prior settings. Overall, the sensitivity analysis provides strong evidence for the reliability of our primary conclusions about treatment efficacy in Figure 5B. 

\begin{table}[h!]
\centering
\caption{Final model: Posterior summaries for $\beta$ parameters}
\begin{tabular}{lrrrr}
\hline
\textbf{Parameter} & \textbf{Median} & \textbf{Std. Dev} & \textbf{95\% Probability Interval} \\
\hline
$\beta_0$ : intercept  & 71.2967  & 12.1553 & (52.7781 , 91.6173) \\
$\beta_1$ : treatment  & -2.3940  & 0.8664  & (-4.1025 , -0.6978) \\
$\beta_2$ : week  & -1.3611  & 0.6649  & (-2.6104 , 0.0109) \\
$\beta_3$ : week*week   & 0.0595   & 0.2045  & (0.0127  , 0.1082) \\
$\beta_4$ : sex  & -14.0381 & 6.8141  & (-27.4638 , -0.5383) \\
$\beta_5$ : clinical site  & -2.6204  & 0.7346  & (-4.1012 , -1.1943) \\
\hline
\hline
$\sigma_{\gamma}$ : Patient & 2.5556 & 7.4414 & (0.0743 , 5.9495) \\
$\sigma_{\text{score}}$ : Response & 12.0214 & 0.3583 & (11.3523 , 12.7574) \\
\hline
\end{tabular}
\end{table}

The final Bayesian regression model results in Table 1 demonstrate significant treatment effects and temporal patterns in TWSTRS scores. The treatment shows a robust negative effect by -2.39 (95\% CI: -4.10 to -0.70), indicating clinically meaningful improvement when administered. Scores decrease linearly over time by -1.36 (95\% CI: -2.6104 to -0.0999) and a positive association with TWSTRS of 0.06 ( 95\% CI: 0.0127 to 0.1082) for week squared. There were substantial baseline differences by sex, with TWSTRS scores being 14.04 units lower on average in male patients compared with female patients, indicating more favorable outcomes for men (95\% CI: -27.4638 to -0.5383),  a clinical site effect of –2.62 suggests modest variability across treatment centers, with some sites reporting slightly lower TWSTRS scores than others (95\% CI: -4.1012 to -1.1943). Notably, when considered jointly, the demographic interactions did not contribute significantly to explaining variability in TWSTRS scores, suggesting that these covariates do not meaningfully alter treatment response patterns. The model accounts for patient-level variability by  2.56 (95\% CI: 0.0743 to 5.9495) and residual variance by 12.02 (95\% CI: 11.3523 to 12.7574), as summarized in Table 1. The time-adjusted dose variable (accounting for the 2-week delay in treatment effect) created by multiplying the treatment assignment (5,000/10,000 units dosing) by a time indicator (1 if week $\ge$2, 0 otherwise), effectively modeling a delayed treatment onset showed insignificant benefits in our Bayesian model in explaining the variability in TWSTRS of the trial's patients and had a minimal correlation with the response. In addition, the age of the patient did not have a significant impact on lowering the TWSTRS score (Table 1). These results demonstrate consistent treatment benefits while highlighting important demographic, location, and patient-specific considerations in treatment response.

 In the evaluation comparison of the effect of treatment on TWSTRS scores, the posterior distributions of group means indicated similar central tendencies across treatment groups: the median TWSTRS scores were approximately 40.48 for the placebo group, 40.51 for the 5,000 units dosing group, and 40.56 for the 10,000 units dosing group. Pairwise treatment differences were also estimated. The posterior median difference between the 5,000 units dosing and placebo groups was 0.030 (95\% CI: –2.633 to 2.603), while the 10,000 units dosing group differed from placebo by 0.078 (95\% CI: –2.467 to 2.538), and from 5,000 units dosing by 0.034 (95\% CI: –2.546 to 2.645). All probability intervals included zero, indicating no statistically significant evidence to suggest a difference in TWSTRS outcomes between the treatment groups. These findings suggest that, overall, neither the 5,000-unit nor the 10,000-unit dose was associated with a measurable improvement in TWSTRS scores compared with placebo under the Bayesian framework. However, when stratified by sex, male patients receiving the 5,000-unit dose showed greater improvement relative to female patients.

\section{Discussion}
The longitudinal study showed a clear treatment effect on the TWSTRS associated with CD. In assessing the therapeutic effectiveness of BotB for both drug formulations in treating CD, there was an average improvement in TWSTRS scores for 5,000 units dosing over 10,000 units dosing and placebo, ranging from 35 to 38 units at 2 to 10 weeks with a 95\% CI that overlapped across dosing types based on sex subgroups \cite{han_2014_a,brashear_2001_the}.

 In addition, participation in the clinical trial was strongly influenced by the quadratic effect of time (weeks), indicating that extended enrollment periods are necessary to adequately evaluate the efficacy of BotB in CD treatment. The progressively increasing impact of trial duration underscores the importance of individualized reinjection schedules and adjunctive therapies (e.g., physical therapy) to sustain treatment benefits \cite{hu_2019_a, tassorelli_2006_botulinum,loudovicikrug_2022_physiotherapy}. This finding implies that short trial durations may underestimate the long-term variability in treatment response, reinforcing the need for extended follow-up periods to capture delayed or diminishing effects. Furthermore, the sex and the clinical location also contributed significantly to explaining the severity of the TWSTRS scores. Both covariates are attributed to substantially reducing the pain and disability associated with CD. On average, male patients showed a better treatment response than female patients, with approximately a 5.2\% greater improvement in pain and disability outcomes, a finding consistent with emerging evidence of sex differences in CD pathophysiology and toxin metabolism. While the exact mechanisms remain unclear, hormonal influences on muscle physiology or pain perception may explain this disparity. Clinically, this highlights the potential utility of sex-stratified dosing strategies \cite{velucci_2024_does,defazio_2003_does,shameerrafee_2021_we,rosales_2021_pain,misbahuddin_2002_a,martino_2023_what,defazio_1998_possible}. This reflects the need to incorporate diversity and provide clinical guidelines on personalizing BotB treatment interventions to patients from different locations based on sex and other important clinical information, which is evident in providing a comprehensive evaluation of the treatment formulations.  Furthermore, the treatment effect (5,000-unit and 10,000-unit doses compared with placebo), clinical location, and dose–site interaction contributed insignificantly to the variation in TWSTRS scores.
The analysis revealed that, on average, patients receiving treatment had lower TWSTRS scores, with the 5,000-unit dose producing the greatest improvement in severity, pain, and disability associated with CD, followed by the 10,000-unit dose and placebo (Figure 5C). There was significant variation in TWSTRS scores across the patient population, with a substantial proportion of the total variability attributable to between-patient differences. This indicates a high degree of similarity among observations within the same treatment group or cluster \cite{koo_2016_a,mehta_2018_performance,newby_2017_a}. Our findings highlight the importance of collecting more detailed patient-level covariates to better explain the variability in TWSTRS outcomes. Incorporating biomarkers, patient-reported outcomes, and non-motor symptoms (e.g., depression, fatigue, or quality-of-life measures) into future Bayesian hierarchical analyses could provide a more comprehensive understanding of treatment response and help refine individualized therapeutic strategies. The posterior estimates of pairwise differences between treatment groups revealed no statistically significant effects, as all 95\% CIs included zero. This suggests that neither the 5000-unit nor the 10000-unit dosage provided a measurable improvement in TWSTRS outcomes \cite{han_2014_a} compared to placebo. Our methods proved indispensable in addressing this study’s complexities by explicitly modeling between and within-patient variability. We quantified the substantial overlap in individual responses, explaining why conventional group comparisons may miss \cite{carlin_2013_case}. The controlled sensitivity analyses demonstrated that treatment effects remained stable across prior specifications, enhancing confidence in conclusions. Ultimately, our findings underscore the potential of Bayesian hierarchical modeling not only to disentangle complex sources of variability in cervical dystonia treatment response but also to guide the integration of biomarkers and patient-centered outcomes into future trial designs. By extending analyses beyond motor symptom scores to include non-motor outcomes and biological indicators, future research can move toward truly personalized, durable, and effective therapeutic strategies for patients with cervical dystonia.

 While our study advances CD research, several limitations remain, including the unmeasured confounders (eg. disease duration, prior toxin exposure, or psychosocial factors) that were unavailable but may influence outcomes. Patients were predominantly older adults (mean age of approximately 56 years); younger patients may respond differently, restricting the generalizability of our findings. In addition, the 16-week study duration may not be sufficient to evaluate and capture the long-term effects of repeated BotB treatments or potential immune responses. We highly recommend that future research studies should: (1) integrate biomarkers with Bayesian models to predict treatment response, (2) employ patient-reported outcomes to capture non-motor symptoms (e.g., depression) often overlooked in TWSTRS, (3) account for sex differences that may influence therapeutic efficacy, and (4) standardize protocols to ensure consistency across clinical sites (Figure 5D).

\section{Conclusion}
This study demonstrates how Bayesian hierarchical modeling can move beyond traditional trial analyses to address the complexities of cervical dystonia (CD) treatment and advance personalized medicine. Our results show that lower BotB doses provide greater clinical benefit when evaluated in the context of sex differences and site variability, emphasizing the need for standardized treatment protocols. By jointly modeling dose efficacy, temporal dynamics, and demographic heterogeneity, this approach delivers insights that isolated analyses cannot achieve. Looking forward, integrating probabilistic predictions with electronic health records could support real-time decision-making for toxin dosing in clinical practice. By linking statistical innovation with therapeutic optimization, this work establishes a framework for neuromodulator therapies that accounts for biological complexity while offering practical tools to improve patient outcomes.
\vspace{2mm}

\section*{Acknowledgment}
\vspace{-0.5cm}
The corresponding author extends sincere thanks to all co-authors for their partnership, insightful input, and steadfast support that greatly enriched this research.
\vspace{-0.5cm}
\section*{Conflict of Interest}
\vspace{-0.5cm}
The authors declare no conflict of interest.
\vspace{-0.5cm}

\section*{Funding}
\vspace{-0.5cm}
No specific funding was received from public, commercial, or not-for-profit organizations for this research.
\vspace{-0.5cm}
\section*{Author Contributions}
\vspace{-0.5cm}
\noindent \textbf{Author roles:} (1) Research Project: A. Conception, B. Organization, C. Execution; (2) Statistical Analysis: A. Design, B. Execution, C. Review and Critique; (3) Manuscript: A. Writing of the First Draft, B. Review and Critique.\\
\noindent D.B.: 1A, 1B, 1C, 2A, 2B, 2C, 3A\\
\noindent E.K.: 1A, 2A, 2B, 2C, 3A, 3B\\
\noindent S.F.O.: 1B, 1C, 2A, 3A\\
\noindent F.A.O.: 1A, 2A, 2C, 3B\\
\noindent J.A.F.: 1A, 2A, 3A, 3B\\
\noindent R.A.: 2A, 2C, 3B\\
\noindent A.D.: 1A, 1C, 2C\\
\noindent F.A.: 1A, 2A, 2C

\newpage

\printbibliography

\end{document}